\documentstyle[preprint,aps]{revtex}
\begin{document}
\def\eps{\epsilon}
\def\up{\uparrow}
\def\down{\downarrow}
\def\be{\begin{equation}}
\def\ee{\end{equation}}
\def\tc{{T_c}}
\def\214{{La_2 Cu 0_4}}

\begin{title} 
{ The ``Spin Gap'' in Cuprate Superconductors}
\end{title}

\author{ Philip W. Anderson}

\address {Joseph Henry Laboratories of Physics\\
Princeton University, Princeton, NJ 08544}
\maketitle

\begin{abstract}
We discuss some generalities about the spin gap in cuprate
superconductors and in detail, how it arises from the interlayer
picture. It can be thought of as spinon (uncharged) pairing,
which occurs independently at each point of the 2D Fermi surface
because of the momentum selection rule on interlayer
superexchange and pair tunneling interactions. Some predictions
can be made.

\end{abstract}
\vfill\eject

The problem with the Spin Gap\cite{1} is that there are too many right
ways to understand it within the interlayer theory\cite{2} 
not too few: when one realizes what is
going on  it seems all too obvious in several ways that one
should have known all along.

(1) The most obvious: spinon pairing. We have realized all along
that the normal state has charge-spin separation, so why didn't
we expect two pairings, one for spin and the second for charge?

(2) Also obvious: there is no phase transition, hardly even a
crossover. So the gap opens without change of symmetry or
condensation. It must be not a self-consistent mean field but a 
property of the separate Fermi surface excitations.

(3) Finally, when one looks at the interlayer theory, and takes
it seriously, one realizes that the phenomenon jumps out at you
and is a trivial consequence of the interlayer interaction.
The Strong-Anderson\cite{3} model is not a complete theory, but can be
used to calculate with: $\chi(T)$, for instance.

Let me start, then, in the inverse of chronological order and try
to make the synthetic argument first. We start from the fact that
every experimental, computational, and theoretical bit of evidence
we have supports the dogma that the 2D interacting electron gas
in the cuprates is a liquid of Fermions with a Fermi surface, and with little or
no tendency to superconductivity or to exhibit
antiferromagnetism, once it is metallic---i.e., there is no
clear indication of ``antiferromagnetic spin fluctuations'', 
as relatively soft bosonic modes, in
the isolated plane. Rather, in the plane the magnetic interaction
modifies the elementary excitation spectrum as it does in the
ferromagnetic case. The symmetry of this state is the
Haldane-Houghton\cite{4} Fermi liquid symmetry 
$(U(2))^Z=(U(1)\times SU(2))^Z$, 
one of ``Z'' for each point on the Fermi surface. 
This large symmetry is the general description of a liquid of
Fermions with a Fermi surface, which is necessarily a surface in
$k$-space on which the Fermion lifetime becomes infinitely long
in the limit as one approaches the surface, hence particles at
the surface are conserved.
Every
point on the Fermi surface is independent, and charge and spin
are separately conserved. The reference shows that this
description includes, but is not confirmed to, the Landau Fermi
liquid. For the Fermi liquid, $U(2)$ applies: the two spin
components are uncoupled; but the basic symmetry is spin and
charge separately conserved, in the general case. 

Our theory$^7$  postulates that in fact the $U(2)$ $\rm\underline { is
}$ broken into $U(1)\times SU(2)$ with the
charge and spin excitations having different Fermi velocities and
the charge also having anomalous dimension, i.e., the charge
bosons are a Luttinger liquid; but this does not change the
symmetry argument. What is little realized is that the spin
excitations are $\rm\underline { always}$ describable as spinons,
even for free electrons, 
$$ \psi^*_{\widehat k} (r)\simeq s^+_{\widehat k} (r)
e^{i\theta_{\widehat k}(r)}$$
The spin part is always a spinon, the charge is  a bosonized
Luttinger liquid.   This, then, is our high-temperature, high
energy state above temperatures and energies where the interplane
interactions come into play. 

Spinons in 2D are paired but gapless.
What the non-existence of a phase transition when we lower $T$ to
the interplanar scale tells us is that the spin gap state has
the same symmetry. It must leave the crucial fact of Fermi or
Luttinger liquids intact: the independence of different Fermi
surface points. Then all that can happen is that the spectrum at
each point changes, and the simplest way for that to happen is
for the spinon to acquire a ``mass'', i.e., the spinons which
used to have a free electron like linear spectrum
$$v_s(k-k_F) \ \ {\rm or}\ \ v_s \sin{\pi\over 2}{(k-k_F)\over k_F}$$
open a gap and have energies 
$$E^2=\Delta^2(\hat k) +v^2_s(k-k_F)^2.\eqno(1)$$ This is possible
because of the peculiar nature of spinons, that they are BCS
quasi-particle like even in the normal state (as shown long ago
by Rokshar\cite{5}). That is, they are
semions, or Majorana Fermions, which have no true antiparticles
(we use the convention $-k=-k,-\sigma\ \ k=k,\sigma$)
$$s^+_k=s_{-k}\qquad s_k=s^+_{-k}$$
so that the Hamiltonian for free spinons may be written
$$v_s(k-k_F)(s^+_ks^+_{-k}+s_{-k}s_k)\eqno(2)$$
just as well as in terms of $s_k^+s_k$
and it is 
$\rm\underline { not \ a \ symmetry \ change }$ to add a term 
$$\Delta_k\ s^+_k\ s^+_{-k}.$$
Spinons are always effectively paired. (Strong and Talstra\cite{6})
It is natural that spinons are more easily paired in the
underdoped regime, because the spinon velocity becomes
progressively lower ($J$ smaller) as we go toward the Mott
insulator; therefore the density of states is higher, $\chi_{\rm
pair}$ larger, on the underdoped side.

Finally, let me make one last remark of a synthetic, rather than
analytic, nature. As I have already said the basic description
either of a Fermi or a Luttinger liquid is the independence of
different Fermi surface points. If we are to go smoothly from a
two-dimensional electron liquid to a gapped state 
$\rm\underline { without \ change\ of \ symmetry}$---without introducing any 
new correlations---we must
do so without coupling the different Fermi surface points, that
is we need 
interactions which conserve two 
dimensional momenta $k_x\  k_y$. There is only one source of such interactions,
namely the interlayer tunneling.
$${\cal H}_{IL}=\sum_{k,\sigma,i,j}\  
t_\perp(k) \ c^+_{ki\sigma} \ c_{kj\sigma}\eqno(3)$$
which, in second order, leads to two types of interlayer
coupling: 

\noindent
Pair tunneling
$${\cal H}_{PT}=\lambda_J(k)\sum_{(ij),k,k'}\ c^+_{k\up i}\ c^+_{-k'\down i}\
c_{-k'\down j}\ c_{k\up j}\eqno(4)$$
and superexchange 
$${\cal H}_{SE}=\lambda_S(k_)\sum_{(ij) k,k'}\ c^+_{k\up i}\ c^+_{-k'\down j}\
c_{-k'\down i}\ c_{k\up j}\eqno(5)$$
(In both, $k'\simeq k$) which represent exchange of charge and spin,
respectively, between two layers. The empirical (and theoretical)
fact that coherent single-particle hopping does not take place in
the cuprates leaves these as the two second-order terms which can
lead to coherent interactions---such as we are looking for---between 
two layers. 

It is important to recognize that (4) and (5) have one extra
conservation relative to conventional interactions. This seems to
be very difficult for many theorists to grasp. 

(5) does not involve any charge exchange between planes hence can
be thought of as an exchange of a spinon pair, if one likes, but
as we shall see it is formally unnecessary to write it in terms
of spinons. (4) only conserves total charge of the two planes,
hence is not a true spinon operator at all. Nonetheless we find
that (4) and (5) together can be described in a sense as pairing
spinon states\cite{7}

This superexchange interaction does not much resemble that used by
Millis and Monien,$^1$ and it does not have anything to do with the
``J'' of the $t-J$ model. Superexchange occurs as a result of
frustrated kinetic energy, and the kinetic energy which is
frustrated in the cuprate layer compounds is only the c-axis
kinetic energy $t_\perp$. They are very like Mott
insulators in one of 3 spatial dimensions: and they exhibit
superexchange in that dimension. But they retain no Mott
character in the 2 dimensions of the planes. 

It is an unpublished conjecture of Baskaran that
$\lambda_S/ \lambda_J$ increases as we approach the insulating
phase, i.e., as ``$\alpha$'', the Fermi surface exponent,
increases. This may be one other reason why underdoped
materials show the spin gap. 

Now, finally, let us do the calculational problem. At this point
we have to stop talking in generalities and make some rather
severe assumptions in order to make progress.
They seem innocuous, and are quite standard in conventional BCS
theory, but here we have no particular reason to believe that they
will serve as better than a rough guide. These assumptions  are:
(1) the Schrieffer pairing condition, i.e., we use only the BCS
reduced interaction $-k'=-k$. This is justified at high enough
$T$ by the fact that a given state $k$ can only pair with one
other $-k'$ to give a quasicoherent matrix element; our
picture of the kind of process involved is that a transition into
a high-energy state intervenes between two low-energy states which
are connected by two---and only two---single-particle 
tunneling processes, $k_a\to
k_b;\ -k_b\to-k_a$. It is perhaps best to think of the pairing as
always $k, -k$ but with center of mass momentum thermally
fluctuating. (2) More orthodox but more serious:
We neglect $|v_c-v_s|$ and treat $c^+_k$ as though it were an
eigenoperation, i.e.

$${\cal H}_K=\sum_k\epsilon_kn_k\eqno(6)$$
Actually we use the Nambu-PWA form
$${\cal H}_K(k)=\eps_k(n_k+n_{-k}-1)\quad = \eps_k\tau_{3k}\ \ .$$
Now we have a straightforward Hamiltonian which is trivially
diagonalized, because it separates into separate Hamiltonians for
every $k$.
$${\cal H}=\sum_k{\cal H}_k$$
$${\cal H}_k={\cal H}_K(k)+\lambda_j\ c^+_{k1}\ c^+_{-k1}\ c_{-k2}\
c_{k2}+1\leftrightarrow 2+\lambda_S\ c^+_{k1}\ c^+_{-k2}\
c^+_{-k1}\ c_{k2}$$
(Here we use the convention $k=k\up$ $-k=-k\down$). The first
attempt was made by Strong and Anderson neglecting $\lambda_s$
and this leads to a beautiful spin gap. The $KE$ spectrum of the 4
fermions 1,2, $k, -k$ has $16=2^4$ states which are grouped into
5 sets, $n_{\rm tot}=0,1,2,3,4$. (See Fig. 1) Of these only the
$n=2$ states are affected by the interactions, and of these 2 
will be split off by $H_J$ and 2 by $H_S$. In either case, 
these gaps are completely $T$-independent and are simply
manifested as the individual states drop out:
$$Z=16\ cosh^4 {\beta\eps_k\over 2}+2\ (cosh\ \beta\lambda_J-1)$$
(because with the added ``-1'' $n=2$ states are at 0 energy.

$\chi$ for this case is 
$$\chi=\int^\infty_{-\infty} d\eps\  {cosh^2\beta \eps/2\over
cosh^4\ \beta\,{\eps\over 2}+{1\over 8}\,(cosh\ \beta\, J-1)}$$
A second calculation may be carried out with both terms,
$\lambda_J\simeq\lambda_S$ and the result is to split out two levels
rather than one and to replace 1/8 with 1/4. This 
is the curve for susceptibility
I show in Fig. 3 and it is not a bad fit to 
susceptibility data. 

But actually I am not totally convinced that this is the right
formalism, although it may be the right arithmetic. The reason it
works seems clearly to me to be that we have picked a form for
the pairing Hamiltonian that connects states which are
``neutral''---i.e., only the $n=2$ states are connected to each
other within the $k$ manifold. But in some real sense these
are states with the spinons paired but with no holon pairing---no
charge pairing---at all, even though nominally different layers
are connected. I think it is more nearly valid to describe the
correct state by rewriting ${\cal H}_j+{\cal H}_s$ as
$$({\cal H}_J+{\cal H}_S)_k \simeq c^+_{ke}\ c^+_{-ke}\ c_{-ke}\ c_{ke}$$
where $c^+_{ke}=
{ c_{k1}+c_{k2}  \over \sqrt{2} }$
That is, the spin-gap state is a state in which spinons belonging
to the $\rm\underline { even }$ linear combination are paired,
 the $\rm\underline { odd }$ unpaired. This
has a strong relationship to the Keimer neutron selection rule
observed for the superconducting state.\cite{8} 
Keimer has begun neutron investigations on spin-gap material, but
his results are completely preliminary. I anticipate that he will
see peaks at energies corresponding to the spin gap and that they will
satisfy his even$\leftrightarrow $odd sum rule, which results from
this pairing.

One consequence of the assumption of Fermi rather than Luttinger
liquid is the $T$-independence of the spin gap. Actually, the
broadening of single-particle states $\propto kT$ will damp out 
the spin gap when $KT>\Delta_{SG}$, as seems to be observed. But
at low $T$, $\Delta_{SG}$ will not vary with $T$. 

This has been a very preliminary account of this work, which is
emphatically in progress. I have benefitted from discussions with
many people, especially Steve Strong, but also T.V. Ramakrishnan,
S. Sarker,  G. Baskaran, D. Clarke; S-D. Liang helped me with the
integral.

\end{document}